\documentclass[prd,nofootinbib,showpacs,showkeys,twocolumn]{revtex4}

\usepackage{graphicx}
\usepackage{amsmath}
\usepackage{amssymb}
\usepackage{amsfonts}
\usepackage{latexsym}
\usepackage{mathrsfs}
\usepackage{color}
\usepackage{slashed}
\usepackage{float}
\usepackage[all]{xy}
\usepackage{hyperref}
\usepackage{dsfont}
\usepackage[justification=centerlast,singlelinecheck=false,font=small]{caption,subfig} 
\global\long\def\norm{\sqrt{-\partial_{\mu}T\partial^{\mu}T}}

\global\long\def\normsq{-\partial_{\mu}T\partial^{\mu}T}

\global\long\def\pa{\partial}

\global\long\def\t{\theta}

\global\long\def\l{\lambda}

\global\long\def\m{\mu}

\global\long\def\n{\nu}

\global\long\def\d{\partial}

\global\long\def\mcM{\mathcal{M}}

\global\long\def\mcT{\mathcal{T}}

\begin{document}


\title{Gauge-Invariant Average of Einstein Equations for finite Volumes}

\author{Juri Smirnov}
\email[\,]{juri.smirnov@mpi-hd.mpg.de}

\affiliation{Max-Planck-Institut f\"ur Kernphysik, Saupfercheckweg 1, 69117 Heidelberg, Germany}

\begin{abstract}
For the study of cosmological backreacktion an avaragng procedure is required. In this work a covariant and gauge invariant averaging formalism for finite volumes will be developed. This averaging will be applied to the scalar parts of Einstein's equations. For this purpose “dust” as a physical laboratory will be coupled to the gravitating system. The goal is to study the deviation from the homogeneous universe and the impact of this deviation on the dynamics of our universe. 
Fields of physical observers are included in the studied system and used to construct a reference frame to perform the averaging without a formal gauge fixing. 
The derived equations resolve the question whether backreaction is gauge dependent.
\end{abstract}

\maketitle

\section{\label{Introduction}Introduction}

Large scale isotropy and homogeneity is the most fundamental assumption of modern cosmology. Those assumptions make it possible to solve Einstein's equations, which reduce to two differential equations with only one parameter, the scale factor. To test this hypothesis and to estimate the effect of the deviations from homogeneity and isotropy an averaging formalism was developed in \cite{Buchert:2001sa,Buchert:1999mc,Buchert:1999er}. It has been critisized in \cite{Gasperini:2009mu, Marozzi:2010qz} that this approach relies on an a priory fixed frame and is thus gauge dependent. An alternative has been suggest which is valid for infinite spacial volumes. In this work we use a physical reference frame, the frame of comoving dust, to perform the averaging in a manifestly gauge independent way. Our formalism is applicable to finite volumes. At the end we compare our result to the equations obtained in \cite{Buchert:1999mc,Buchert:1999er} and show that they are equivalent in the frame of the introduced pressure-less fluid.

\section{\label{Gauge invariant Buchert equations} Gauge invariant Buchert equations}

Starting from local inhomogeneous Einstein equations an averaging procedure is set up to derive effective equations which quantitatively measure the deviation from the Friedmann equations which describe the dynamics of a homogeneous and isotropic space-time. To avoid the question of coordinate fixing and hence gauge dependence a system of freely falling observers will be introduced and used to perform the averaging. It is important to mention
the compatibility of the dust as coordinate system with a deformed
FRW universe, as it respects the fundamental symmetries of the background. 
We know that our universe used to be homogeneous to a very high precision and hence was described
well by the FRW universe. However, nowadays there are large inhomogeneities
locally present in the universe. Nevertheless, observations of the large scale structure suggest that the underlying geometry is not drastically different from an homogeneous and isotropic space on horizon scales. Therefore,
it is sensible to use a laboratory which does not destroy the FRW
space-time, since it has the appropriate symmetries. The dust we are
going to use fulfils this criterion.

\subsubsection*{\textbf{The model}}

We consider a space-time filled with matter and demand that the energy
density of this matter should never vanish. This assumption allows to avoid the problem of seemingly indeterministic behaviour of General Relativity (GR), known as the hole problem \cite{Rovelli:2004tv}. Demanding a non vanishing matter content we are able to give physical meaning to every point
in space-time. This is also a reasonable assumption on physical grounds,
since even the voids contain some amounts of cosmic dust or radiation.
Furthermore, it is speculated that there is matter present in the universe,
which interacts only (or almost only) with gravity. The dark matter
is considered to be ``cold'' i.e. non- relativistic in the standard
model of cosmology, hence it can be well described by dust. Thus, our model can be viewed as a 
realistic cosmological model incorporating the cold dark matter.

The distribution of matter in our model though is not homogeneous,
since it is supposed to represent a universe perturbed about the FRW background symmetry. This is
the difference to the approach in \cite{Gasperini:2009mu, Marozzi:2010qz}, we allow for matter
to exist which is distributed inhomogeneously and hence it is possible
to find physical fields with a space-like gradient. 

Within our model the energy momentum tensor can be decomposed in the
way described below, since we have demanded for the energy density
never to vanish. We can view this in two ways. Either the dust of
low energy is formally separated from the total energy momentum tensor,
or it is added to it, which would not affect the system strongly.

\begin{align}
T^{\mu\nu}=T_{\textrm{dust}}^{\mu\nu}+\tilde{T}^{\mu\nu}\,,
\end{align}

where the $T_{\textrm{dust}}^{\mu\nu}=\epsilon\, u^{\mu}u^{\nu}$
is the dust energy momentum with infinitesimal $\epsilon$ and $\tilde{T}^{\mu\nu}$
is the energy momentum tensor of the rest of the system.

Consequently the Lagrangian of the theory is as follows

\begin{align}
L=L_{\textrm{gravity}}+\tilde{L}+L_{\textrm{dust}}\,.
\end{align}

The gravitational Lagrangian is the Ricci scalar multiplied by the
square root of the metric determinant. The dust Lagrangian is  

\begin{align}
L=\sqrt{-g}\,\epsilon\left\{ v^{\mu}v^{\nu}g_{\mu\nu}+1\right\} \,.
\end{align}

With the following eight fields, $\epsilon$ being the energy density,
$T$ eigentime , $W_{l}$ velocity field and $Z^{k}$ coordinate field,
where $k,l\in\left\{ 1,2,3\right\} $ 

\begin{align}
v^{\m}=-\d^{\m}T+W_{k}\d^{\m}Z^{k}\,,
\end{align}

$T$ has a time-like gradient and the $Z^{k}$ 's have obviously space-like
gradients. This fields will be used for the construction of a window
function which transforms as a scalar. We will restrict our analysis
to the Lagrangian formalism and not be concerned with the form of
the physical Hamiltonian. On the other hand we know that a physical
Hamiltonian exists, since the system with dust deparametrizes, as
has been shown in \cite{Dittrich:2005kc,Giesel:2007wk,Thiemann:2006up}.

\vspace{3mm}

\subsection{Time}

The eigentime of the dust particles is a map into the dust-time manifold

\begin{align}
T:(\mathbb{R}\times\Sigma)\rightarrow\mcT\,.
\end{align}

At each fixed value $T=\tau$ , $T$ is a
scalar w.r.t. coordinate changes on the manifold $\mcM$ in the sense

\begin{align}
\tilde{T}(X)=T(\tilde{X})\,,
\end{align}

where $\tilde{X}=f^{-1}(X)$. On the dust-time manifold $T$ is trivially
a scalar. This will be scalar with a space-like gradient which will
be used for the construction of the window function. We take the eigentime
$T$ of the dust particles as the scalar field which generates the
time flow. Therefore the unit time-like vector is

\begin{align}
n^{\mu}=\frac{-\partial^{\mu}T}{\sqrt{-\partial_{\mu}T\partial^{\mu}T}}\,.
\end{align}

The foliation is defined as the hypersurfaces on which $T$ takes
constant values. It is the physical time of the dust field. This defines
a time-space split with the projector orthogonal to the time flow

\begin{align}
h_{\mu\nu}=g_{\mu\nu}+n_{\mu}n_{\nu}\,.
\end{align}

\vspace{3mm}

\subsection{Space}

The dust is a collection of geodesic observers, freely falling on
$\mcM$. Each particle carries a label, at any instant of time $\tau$
a particle has a position, so this position of the labeled particle
is a physical value and hence a Dirac observable. Taking the continuum
limit, while keeping the particle density constant we obtain a field,
where the indices of the particles can be viewed now as three dimensional
coordinates $Z^{i}\,,\, i\,\in\left\{ 1,2,3\right\} $. $Z$ is a
map from the hypersurface of constant time into the dust-space manifold
of constant $T=\tau$

\begin{align}
Z:\,\Sigma\rightarrow S(\tau)\,.
\end{align}

Now we define the quantity: $\bar{Z}^{i}(T(\tau)):=Z^{i}(\tau)$ . The $\bar{Z}^{i}$
are for evry fixed pair of values $Z^{i}$ and $T$ scalars on $\mcM$,
(omit the bar in the coming discussion)

\begin{align}
\tilde{Z}^{k}(X)=Z^{k}(\tilde{X})\,,
\end{align}

where $\tilde{X}=f^{-1}(X)$. On the other hand they are three vectors
on $S$. To construct a quantity which is a scalar on $\mcM$ and
$S$, more work is required. The basic idea is to construct a scalar
on $S$ out of the coordinate fields of the dust, a quantity like
$Z_{k}Z^{k}$. Therefore we need the induced metric on the hypersurfaces
$h_{ij}$. To get this object we pull back the orthogonal projector
$h_{\mu\nu}$ on the hypersurface of constant $T$ denoted by $\Sigma$.
To stay general we introduce coordinates on $\Sigma$ called $x^{a}$
which represent a map 

\begin{align}
Z^{-1}:\, S(\tau)\rightarrow\Sigma\,.
\end{align}

Furthermore, the coordinates on $\mcM$ are denoted by $X^{\m}(x^{a})$
and represent a map 

\begin{align}
X:\,\Sigma\,\rightarrow\mcM\,.
\end{align}

The combined map which we are going to use for the pull-back is $\rho$,
defined as

\begin{align}
\rho:=X\circ Z^{-1}\;\Rightarrow\rho:\, S(\tau)\rightarrow\mcM\,.
\end{align}

We can visualize the interrelations as follows

\begin{figure}[H]
\begin{centering}
\includegraphics[scale=0.5]{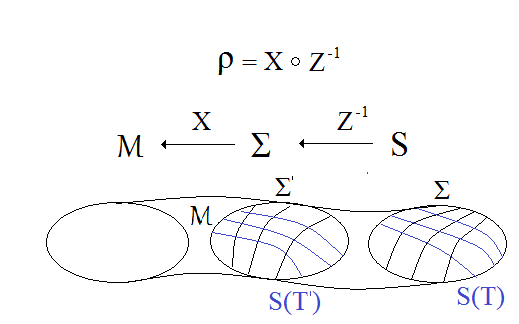}
\par\end{centering}

\caption{This pictogram shows the falling of the geodesic observes on the manifold $\mathcal{M}$, the hypersurfaces of constant dust-time are denoted by $\Sigma$ and $\Sigma'$. While the map $Z^{-1}$ goes from the dust-space to the space of internal coordinates of the hypersurface of constant dust-time, the map $X$ goes from this internal space to the manifold. Therefore, the combination of both maps goes from the dust-space slice to the full space-time manifold and is used for the pull-back.}
\end{figure}

With the pull backs of the maps given as

\begin{align}
X_{*}=\frac{\d X^{\m}}{\d x^{a}}=:X_{a}^{\m}\,,
\end{align}

\begin{align}
Z_{*}^{-1}=\frac{\d x^{i}}{\d Z^{j}}=:(Z^{-1})_{j}^{a}\,,
\end{align}

\begin{align}
\rho_{*}=\frac{\d X^{\m}}{\d Z^{j}}=X_{a}^{\m}(Z^{-1})_{j}^{a}\,.
\end{align}

The choice of the $X$ and $Z^{-1}$ maps corresponds to the gauge
choice. Later we will evaluate the result in the the so called ADM gauge \footnote{The Arnowit, Deser, Misner (ADM) formalism was developed to study the Hamiltonian evolution of a gravitating system \cite{Arnowitt:1962hi}.} where $\rho=X$
and $Z^{-1}= \mathbb{I}$, therefore $\Sigma=S$. After defining the maps
we come back to the main goal and pull-back on the dust space

\begin{align}
h_{ij}(Z) &  =\rho_{*}h_{\m\n}=X_{a}^{\m}(Z^{-1})_{i}^{a}X_{b}^{\n}(Z^{-1})_{j}^{b}\, h_{\m\n}= \\ \nonumber
& = \frac{\partial X^{\mu}}{\partial Z^{i}}\frac{\partial X^{\nu}}{\partial Z^{j}}h_{\mu\nu}=:X_{i}^{\mu}X_{j}^{\nu}h_{\mu\nu}=X_{i}^{\mu}X_{j}^{\nu}g_{\mu\nu} \\ \nonumber
+  X_{i}^{\mu}& X_{j}^{\nu}n_{\mu}n_{\nu}  =X_{i}^{\mu}X_{j}^{\nu}g_{\mu\nu}=X_{a}^{\m}(Z^{-1})_{i}^{a}X_{b}^{\n}(Z^{-1})_{j}^{b}\, g_{\m\n}\,.
\end{align}

Here the property has been used that $n^{\mu}$ is orthogonal to the
tangents of $S$ : $\partial_{i}X^{\mu}$ , i.e. $X_{i}^{\mu}n_{\mu}=0\;\forall\, i$
.

Note: For the construction we have used the orthogonality of the time
flow to the spatial hypersurfaces. To ensure this we can pick the
dust velocity field $W_{k}$ to vanish. This is simply assuming that
the dust we consider is rotation free, which is reasonable for all
practical purposes.

The constructed quantity commutes with the 3-diffeomorphism constraints and in
the next step we construct the
observable, using the dynamic equation with $t$ being $X^{0}$ from
the ADM split: 

\begin{align}
\frac{\d h_{ij}(Z)}{\d t}=\left\{ h_{ij}(Z)\,,\, H(f)\right\}\,,
\end{align}

here $H(f)=\int d^{3}x\, H_{0}(x)f(x)$ with $f$ some scalar density
and $H_{0}$ the total Hamiltonian constraint of the system. Even if an analytic solution to this equation can
not be obtained, we can set initial values by field re-definition
and perform a series expansion. The solution of this equation $h_{ij}(Z,t)$
is used to define the observable $\bar{h}_{ij}(Z,T)$

\begin{align}
\bar{h}_{ij}(Z\,,\, T(Z,t)):=h_{ij}(Z,t)\,.
\end{align}

This quantity commutes for fixed $T$ and $Z$ with all constraints
and is hence an observable. In the
following we will omit the bar for simplicity. Now we found the
tensor quantity on $S$ which is needed for the contraction with the
vectors and can write

\begin{align}
Z^{2}=Z^{k}h_{kl}\, Z^{l}=:B\,.
\end{align}

 The dust vectors are contracted with a tensor on the dust manifold
and one obtains a physical scalar on $\mcM$ and on $S$. It transforms
as 

\begin{align}
\tilde{B}(X)=B(\tilde{X})\,,
\end{align}

and

\begin{align}
\tilde{B}(x,T)=B(\tilde{x},T)\,,
\end{align}

where on $\mcM$: $\tilde{X}=f^{-1}(X)$ and correspondingly on $S$:
$\tilde{x}=f^{-1}(x)$.

We check the transformation properties below.

\begin{itemize}
\item The induced metric on the hypersurface $X_{a}^{\m}(Z^{-1})_{i}^{a}X_{b}^{\n}(Z^{-1})_{j}^{b}\, g_{\m\n}$
is a scalar under coordinate changes on the manifold by construction,
an explicit calculation shows it in components:

\begin{align}
\tilde{h}_{ij}(X) & =\frac{\pa}{\pa Z^{i}}\frac{\partial\tilde{X}^{\m}}{\partial X^{\lambda}}X^{\l}\frac{\pa X^{\l}}{\pa\tilde{X}^{\m}}\frac{\pa}{\pa Z^{j}}\frac{\partial\tilde{X}^{\n}}{\pa X^{\rho}}X^{\rho}\frac{\pa X^{\rho}}{\pa\tilde{X}^{\n}}g_{\l\rho} \\ \nonumber
& =\frac{\pa X^{\l}}{\pa Z^{i}}\frac{\pa X^{\rho}}{\pa Z^{j}}g_{\l\rho}=h_{ij}(\tilde{X})\,.
\end{align}

\item The constructed quantity $B=Z^{2}$ is a scalar on $S$ since the
3-vectors $Z^{k}$ are contracted with a rank two tensor on $S$.
Again explicit calculation shows:
\begin{align}
& B(\vec{Z})=Z^{k}h_{kl\,}Z^{l}=\left(\frac{\d Z^{k}}{\d x^{a}}x^{a}\right)\tilde{h}_{kl}\left(\frac{\d Z^{l}}{\d x^{b}}x^{b}\right) = \\ \nonumber
& =\left(\frac{\d Z^{k}}{\d x^{a}}x^{a}\right)\left(h_{ab}\frac{\d x^{a}}{\d Z^{k}}\frac{\d x^{b}}{\d Z^{l}}\right)\left(\frac{\d Z^{l}}{\d x^{b}}x^{b}\right)=x^{a}\, h_{ab}\, x^{b}=B(\vec{x})\,.
\end{align}

\item The connecting relation which makes the whole object invariant under
coordinate transformations on the manifold is that the $Z^{k}$ s
are the dust labels and hence scalars on $\mcM$ as discussed above:

\begin{align}
\tilde{Z}^{k}(X)=Z^{k}(\tilde{X})\,,
\end{align}

 where $\tilde{X}=f^{-1}(X)$. 

\end{itemize}

Now we have shown that $B=Z^{2}$ is a scalar on $\mcM$ and on the
dust manifold $S$. Since the $Z^{2}$ is a scalar on $S$ we can
express it in arbitrary coordinates $x^{a}$ on $\Sigma$, the ADM
split hypersurface of constant $T$. This coordinate representation
is also convenient for the Window function which will be discussed in the following
section.

\vspace{3mm}

\subsection{Window function}

Having constructed physical scalars we write the window function
as

\begin{align}
\label{eqn:Window}
W_{\Omega}=\delta(T(x)-T_{0})\:\sqrt{-\partial_{\mu}T\partial^{\mu}T}\,\Theta(r_{0}-B(x))\,,
\end{align}

with $\delta(x)$ is a Dirac delta function and $\Theta(x)$ the step function. The Window function constructed in (\ref{eqn:Window}) is a scalar on $\mcM$ and $S$ i.e. it transforms as 

\begin{align}
W\rightarrow\tilde{W}(x)=W(f^{-1}(x))\label{eq:windowtrafo}\,.
\end{align}

Now the functional is set up in the following way

%

\begin{align}
\label{eqn:Functional}
& F(\Omega)=\\  \nonumber
& \int_{\mathcal{M}_{4}}\sqrt{-g(x)}\:\sqrt{-\partial_{\mu}T\partial^{\mu}T}\,\Theta(r_{0}-B(x))\delta(T_{0}-T(x))d^{4}x\,.
\end{align}

The averaging functional of a quantity $S$ can be defined using Eqn. (\ref{eqn:Functional}), and
simplified integrating out the delta function leading to

\begin{widetext}
\begin{align}
 \left\langle S\right\rangle {}_{\{A,r_{0}\}} & =\frac{F(S,\Omega)}{F(1,\Omega)}=\frac{\int_{\mathcal{M}_{4}}\sqrt{-g(x)}\:\sqrt{-\partial_{\mu}T\partial^{\mu}T}\,S(x)\,\Theta(r_{0}-B(x))\delta(T_{0}-T(x))d^{4}x}{\int_{\mathcal{M}_{4}}\sqrt{-g(x)}\:\sqrt{-\partial_{\mu}T\partial^{\mu}T}\Theta(r_{0}-B(x))\delta(T_{0}-T(x))d^{4}x}= \\ \nonumber
& =\frac{\int_{\Sigma_{0}}\int_{\mathfrak{\mathbb{R}}}N\,\sqrt{h(\vec{x})}\:\sqrt{-\partial_{\mu}T\partial^{\mu}T}S(x)\Theta(r_{0}-B(x))\delta(T_{0}-T(x))d^{3}x\, dT}{\int_{\Sigma_{0}}\int_{\mathfrak{\mathbb{R}}}N\,\sqrt{h(\vec{x})}\:\sqrt{-\partial_{\mu}T\partial^{\mu}T}\Theta(r_{0}-B(x))\delta(T_{0}-T(x))d^{3}x\, dT}= \\  \nonumber
& =\frac{\int_{\Sigma_{0}}N\,\sqrt{h(x)}\:\sqrt{-\partial_{\mu}T\partial^{\mu}T}|_{T=T_{0}}S(x)\Theta(r_{0}-B(x))d^{3}x}{\int_{\Sigma_{0}}N\,\sqrt{h(x)}\:\sqrt{-\partial_{\mu}T\partial^{\mu}T}|_{T=T_{0}}\Theta(r_{0}-B(x))d^{3}x} \,,
\end{align}
\end{widetext}

where $\Sigma_{0}$ is the hypersurface on which $T(x)=T_{0}$ , $h$
the determinant of the induced spatial metric discussed above and
$N$ the lapse function for a general not synchronous gauge. The average
is manifestly a gauge independent quantity, since $T(x)$ and $B(x)$
are scalars under Diff($\mathcal{M}$) and also Diff($S$) and so
the window function is also a scalar, leading to the following transformation property of the averaging functional

\begin{align}
& F(S,\Omega)\rightarrow\tilde{F}(\tilde{S},\Omega)=\int_{\mathcal{M}_{4}}\sqrt{-\tilde{g}(x)}\tilde{S}(x)\tilde{W}_{\Omega}(x)d^{4}x \\ \nonumber
& =\int_{\mathcal{M}_{4}}\left|\frac{\partial x}{\partial f}\right|{}_{f^{-1}(x)}\sqrt{-g(f^{-1}(x))}S(f^{-1}(x))W_{\Omega}(f^{-1}(x))d^{4}x \, ,
\end{align}

introducing $\hat{x}=f^{-1}(x)$ we obtain
\begin{align}
\tilde{F}(\tilde{S},\Omega)=\int_{\mathcal{M}_{4}}\sqrt{-g(\hat{x})}S(\hat{x})W_{\Omega}(\hat{x})d^{4}\hat{x}=F(S,\Omega)\,.
\end{align}

The gauge functional is obviously gauge invariant and hence the average of $S$ is an observable. 

Equipped with this functional we can address the problem of averaging
the scalar parts of Einstein's equations of the model of the universe
presented above, where the coordinate system, the dust, is included
into the studied system. We will perform a similar analysis as in
\cite{Gasperini:2009mu}, but in a more general way, since our formalism applies
also to finite volumes.

\vspace{3mm}

\subsection{Time derivative of the average}

To average the dynamic equations of our model it is also necessary
to compute time derivatives of the averaged quantities. In this case
the time derivative means partial derivative w.r.t $T_{0}$ the eigentime
of the homogeneous dust used for the deparametrisation. Let us begin
with the derivative of $F(S,\Omega)$. The strategy will be to choose
certain coordinates in which the computation will be easier and then
to generalize our result to an arbitrary frame.

The eigentime $T$ is the function that defines the space as hypersurfaces
on which it assumes a constant value, therefore, it is spatially homogeneous.
In this case we can choose coordinates, in future referred to as ADM
coordinates, such that: $n_{\mu}=N\,(-1,0,0,0)$ and $n^{\mu}=\frac{1}{N}(1,\ -N^{i})$
where $N$ is the lapse and $N^{i}$ the shift. The metric reads:

\begin{align}
ds^2=-N^2\,dt^2+h_{ij}(dx^{i}+N^{i}dt)(dx^{j}+N^{j}dt)\,.
\end{align}

Furthermore, since we have set the velocity field $W^{k}$ of the coordinate
dust to zero, which implies the orthogonality of $n^{\mu}$ to the
spatial hypersurfaces, the shift vectors vanish. In this coordinates
we compute the time derivative. The partial derivative commutes past
the integral and the derivative of the $\delta$ distribution we understand
in the weak sense
\begin{widetext}
\begin{align}
\label{eqn:DerivativeAverage}
\frac{\partial F(S,\Omega)}{\partial T_{0}} & =-\int_{\mathcal{M}_{4}}\sqrt{-g(x)}\:\sqrt{-\partial_{\mu}T\partial^{\mu}T}S(x)\Theta(r_{0}-B(x))\frac{\partial}{\partial T}\{\delta(T_{0}-T(x))\}d^{4}x \\ \nonumber
& = -\int_{\mathcal{M}_{4}}\sqrt{-g(x)}\:\sqrt{-\partial_{\mu}T\partial^{\mu}T}S(x)\Theta(r_{0}-B(x))(\partial_{0}T)^{-1}\partial_{0}\{\delta(T_{0}-T(x))\}d^{4}x \,.
\end{align}
\end{widetext}

In the coordinates chosen we can read of from the form of $n_{\mu}$
the only component of $\partial_{\mu}T$ which is non-zero: $\partial_{0}T$
and $-\partial_{\mu}Tg^{\mu\nu}\partial_{\nu}T=-(\pa_{0}T)^{2}g^{00}$
with $g^{00}=-N^{-2}$. Thus, Eqn. (\ref{eqn:DerivativeAverage}) simplifies to

\begin{widetext}
\begin{align}
\frac{\partial F(S,\Omega)}{\partial T_{0}} & = -\int_{\mathcal{M}_{4}}\sqrt{-g(x)}\: S(x)\Theta(r_{0}-B(x))\sqrt{-g^{00}}\partial_{0}\{\delta(T_{0}-T(x))\}d^{4}x  \\ \nonumber
& = \int_{\mathcal{M}_{4}}\partial_{0}\left\{ \sqrt{-g(x)}\: S(x)\Theta(r_{0}-B(x))\sqrt{-g^{00}}\right\} \delta(T_{0}-T(x))d^{4}x \\ \nonumber
& =\int_{\mathcal{M}_{4}}\sqrt{|h|}\left\{ \:\Theta(r_{0}-B(x))\left(N\theta S+\partial_{0}S\right)-\delta(r_{0}-B(x))S\partial_{0}B(x)\right\} \delta(T_{0}-T(x))d^{4}x \,.
\end{align}
\end{widetext}

Here $h=det(h_{ij})$ and $\theta=\frac{1}{N}\partial_{0}\log(\sqrt{h})$,
since $\theta_{ij}=\frac{1}{2}\mathcal{L}_{n}h_{ij}=\partial_{T}h_{ij}=\frac{1}{N}\partial_{0}h_{ij}$
in this gauge and the trace $\theta=\frac{1}{N}h^{ij}\dot{h}_{ij}=\frac{1}{N}\partial_{0}\log(\sqrt{h})$
. Having this, we can rewrite the expression in a covariant way, so
that it reduces to our result when the ADM coordinates are fixed 

\begin{widetext}
\begin{align}
\frac{\partial F(S,\Omega)}{\partial T_{0}}=F\left(\frac{\partial_{\mu}T\partial^{\mu}S}{\norm},\Omega\right)+F\left(\frac{S\theta}{\norm},\Omega\right)-2\, F\left(\frac{\partial_{\mu}T\partial^{\mu}B}{\norm}\: S\,\delta(r_{0}-B),\Omega\right)\,.
\end{align}
\end{widetext}

The last term vanishes in case $n^{\mu}\partial_{\mu}B=0$ i.e. the
spatial coordinates do not depend on the time variable. This is indeed
the case in our model since $n^{\mu}$ is orthogonal to the space
such that $\nabla_{n}Z^{k}=0$ which implies the above relation. Hence
the time derivative reads

\begin{equation}
\frac{\partial F(S,\Omega)}{\partial T_{0}}=F\left(\frac{\partial_{\mu}T\partial^{\mu}S}{\norm},\Omega\right)+F\left(\frac{S\theta}{\norm},\Omega\right)\,.
\label{eq:derivative of functional}
\end{equation}

Now we can calculate the time derivative of the average of a scalar

\begin{widetext}
\begin{align}
 \frac{\partial\left\langle S,\Omega\right\rangle }{\partial T_{0}} & =\frac{\partial F(S,\Omega)}{\partial T_{0}}\frac{1}{F(1,\Omega)}-\left\langle S,\Omega\right\rangle \frac{1}{F(S,\Omega)}\frac{\partial F(1,\Omega)}{\partial T_{0}}= \\ \nonumber
& =\left(F(\partial_{T_{0}}S,\Omega)+F(\frac{N\, S\,\theta}{\partial_{0}T},\Omega)\right)\frac{1}{F(1,\Omega)}-\left\langle S,\Omega\right\rangle \frac{1}{F(1,\Omega)}F\left(\frac{N\,\theta}{\partial_{0}T},\Omega \right)= \\ \nonumber
& =\left\langle \partial_{T_{0}}S,\Omega\right\rangle +\left\langle \frac{N\, S\,\theta}{\partial_{0}T},\Omega\right\rangle -\left\langle S,\Omega\right\rangle \left\langle \frac{N\,\theta}{\partial_{0}T},\Omega\right\rangle \,.
\end{align}
\end{widetext}

Covariantly expressed we obtain the general form of the Buchert, Ehlers
 commutation rule, \cite{Buchert:1999er}:

\begin{widetext}
\begin{align}
\frac{\partial\left\langle S,\Omega\right\rangle }{\partial T_{0}}=\left\langle \frac{\partial_{\mu}T\partial^{\mu}S}{\partial_{\mu}T\partial^{\mu}T},\Omega\right\rangle +\left\langle \frac{S\,\theta}{\norm},\Omega\right\rangle -\left\langle S,\Omega\right\rangle \left\langle \frac{\theta}{\norm},\Omega\right\rangle \,.
\label{eq:commutation rule}
\end{align}
\end{widetext}

\vspace{3mm}

\subsection{The effective scale factor}

In order to be interpreted as a domain dependent scale factor analogous
to the $a$ of the FRW model, a quantity has to be considered which
is gauge independent. As any physical observable must not depend on
the coordinates chosen. 

To accomplish this we use the gauge independent averaging functional
to define the quantity $s$ to be the effective scale factor

\begin{align}
\frac{1}{s}\frac{\partial s}{\partial T_{0}}:=\frac{1}{3}\frac{1}{F(1,\Omega)}\frac{\partial F(1,\Omega)}{\partial T_{0}}\,.
\end{align}

Using (\ref{eq:derivative of functional}) one sees that 
\begin{align}
\frac{\partial F(1,\Omega)}{\partial T_{0}}=F\left(\frac{\theta}{\norm},\Omega\right)\,,
\end{align}

 and hence

\begin{align}
\frac{1}{s}\frac{\partial s}{\partial T_{0}}=\frac{1}{3}\left\langle \frac{\theta}{\norm}\right\rangle \label{eq:adot}\,.
\end{align}

\subsubsection*{Remark:}

A more pictorial way to describe this, is to write 

\begin{align}
s=\left(\frac{F(1,\Omega)_{T_{0}}}{F(1,\Omega)_{0}}\right)^{\frac{1}{3}}
\end{align}

leading to

\begin{align}
\frac{\partial s}{\partial T_{0}}=\frac{1}{3}(s^3)^{-\frac{2}{3}}\frac{1}{F_{0}}\frac{\partial F(1,\Omega)_{T_{0}}}{\partial T_{0}}=\frac{1}{3}s\left\langle \frac{\theta}{\norm}\right\rangle \,.
\end{align}

We see from this calculation that $s$ has indeed a property of a
scale factor, but its definition above circumvent the necessity of
introducing a normalization at a certain initial time point.

\section{Einstein's equations }

We will derive the expressions for the averaged Hamilton constraint density and the Raychaudhuri equation, the important point of the 
derivations is that in our set up the $n^{\m}$,
which defines the time flow of the coordinate dust, is already
present in the model and is not chosen arbitrarily. The time arises
natural and there is no need of artificial gauge fixing. Thus, we describe
the time flow without formally choosing a gauge.

\vspace{3mm}

\subsection{Gauge independent averaging }

\subsubsection*{Normal projection of the equations of motion:}

The Einstein tensor multiplied with the time flow vectors (which is
equivalent to the Hamiltonian constraint density) gives

\begin{align}
G^{\mu\nu}n_{\mu}n_{\nu}=\frac{1}{2}R_{s}+\frac{1}{3}\theta^2-\sigma^2\,,
\end{align}

where the expansion rate and the shear are defined w.r.t $n_{\mu}$,
the $R_{s}$ is the spatial curvature associated with the projector
on the hypersurface orthogonal to $n_{\mu}$.

On the other hand one has from the matter contribution

\begin{align}
& T^{\mu\nu}n_{\mu}n_{\nu}=(T_{\textrm{dust}}^{\mu\nu}+\tilde{T}^{\mu\nu})n_{\mu}n_{\nu} \\ \nonumber
&T_{\textrm{dust}}^{\mu\nu}n_{\mu}n_{\nu}=\epsilon(v_{\mu}n^{\mu})^2=\epsilon(-\partial_{\mu}Tn^{\mu})^2=\epsilon(-\partial_{\mu}T\partial^{\mu}T)\,,
\end{align}

with $v^{\mu}=\norm\: n^{\mu}$ since we said that the velocity field
$W^{k}$ vanishes. Now the Hamiltonian constraint density can be averaged.
For normalization divide it first by $-3\,\partial_{\mu}T\partial^{\mu}T$
apply the averaging functional introduced above and insert the identity
by adding and subtracting the average of $\t$ squared we get 

\begin{widetext}
\begin{align}
\label{eq:hamilton constraint}
\left\langle \frac{1}{6}\frac{R_{s}}{\normsq}+\frac{1}{9}\frac{\theta^2}{\normsq}-\frac{1}{3}\frac{\sigma^2}{\normsq}\right\rangle +\frac{1}{9}\left\langle \frac{\theta}{\norm}\right\rangle ^{2}-\frac{1}{9}\left\langle \frac{\theta}{\norm}\right\rangle ^{2}  
=\frac{8\pi G_{N}}{3}\left\langle \frac{\tilde{T}^{\mu\nu}n_{\mu}n_{\nu}}{\normsq}+\epsilon\right\rangle \,,
\end{align}
\end{widetext}

using

\begin{align}
\left(\frac{1}{s}\frac{\partial s}{\partial T_{0}}\right)^{2}=\frac{1}{9}\left\langle \frac{\theta}{\norm}\right\rangle ^{2} \,,
\end{align}

and defining the backreaction

\begin{align}
Q_{D} & =\frac{1}{3}\left(\left\langle \frac{\theta^2}{\normsq}\right\rangle -\left\langle \frac{\theta}{\norm}\right\rangle ^{2}\right) \\ \nonumber
& -2\left\langle \frac{\sigma}{\normsq}\right\rangle \,,
\end{align}

we rewrite (\ref{eq:hamilton constraint}) as

\begin{align}
\left(\frac{1}{s}\frac{\partial s}{\partial T_{0}}\right)^{2}=-\frac{1}{6}\,\left\langle \frac{R_{s}}{\normsq}\right\rangle -\frac{1}{6}Q_{D} \\ \nonumber
+ \frac{8\pi G_{N}}{3}\left\langle \frac{\tilde{T}^{\mu\nu}n_{\mu}n_{\nu}}{\normsq}+\epsilon\right\rangle \,.
\end{align}

\vspace{3mm}This is the effective equation describing the velocity
of the expansion of the domain dependent scale factor. It is the analogue
to the first Friedmann equation. With the definitions comparable to
those in \cite{Buchert:1999er} we can write the cosmological energy balance:

\begin{align}
H_{D}:=\frac{1}{s}\frac{\partial s}{\partial T_{0}}\,,
\end{align}

\begin{align}
\Omega_{m}:=\frac{8\pi G_{N}}{3\, H_{D}^{2}}\left\langle \frac{\tilde{T}^{\mu\nu}n_{\mu}n_{\nu}}{\normsq}\right\rangle \,,
\end{align}

\begin{align}
\Omega_{\epsilon}:=\frac{8\pi G_{N}}{3\, H_{D}^{2}}\,\left\langle \epsilon\right\rangle \,,
\end{align}

\begin{align}
\Omega_{k}:=-\frac{\left\langle R\right\rangle }{6H_{D}^{2}}\,,
\end{align}

\begin{align}
\Omega_{Q}:=-\frac{Q_{D}}{6H_{D}^{2}}\,,
\end{align}

\begin{align}
\Omega_{k}+\Omega_{Q}+\Omega_{\epsilon}+\Omega_{m}=1\,.
\end{align}

From the energy balance equation we see that the backreaction term
$\Omega_{Q}$ as well as the coordinate mass term $\Omega_{\epsilon}$
enter the balance and the backreaction term opens the possibility
of contributing to an effective cosmological constant term.

\vspace{3mm}

\subsubsection*{Raychaudhuri's equation:}

Combining the Hamiltonian constraint density with the trace of the
Einstein equations projected orthogonal to the time flow, one gets%

\begin{align}
R_{\mu\nu}n^{\mu}n^{\nu}=T_{\mu\nu}h^{\mu\nu}-\frac{1}{2}T\,,
\end{align}

which is equivalent to

\begin{align}
-n^{\mu}\nabla_{\mu}\theta=2\sigma^2+\frac{1}{3}\theta^2-\nabla^{\nu}(n^{\mu}\nabla_{\mu}n_{\nu}) \\ \nonumber
+(T_{\mu\nu}-\frac{1}{2}g_{\mu\nu}T)n^{\mu}n^{\nu}\,.
\label{eq:Rauchaudhuri}
\end{align}

In our model the last term gives

\begin{align}
(T_{\mu\nu}-\frac{1}{2}g_{\mu\nu}T)n^{\mu}n^{\nu} & =T_{\mu\nu}^{dust}n^{\mu}n^{\nu}+\tilde{T}_{\mu\nu}n^{\mu}n^{\nu}+\frac{1}{2}T \nonumber \\ 
 =\tilde{T}_{\mu\nu}n^{\mu}n^{\nu}+\frac{1}{2}\tilde{T} & +\epsilon(-\partial_{\mu}T\partial^{\mu}T)+\frac{1}{2}(-\epsilon)\,.
\end{align}

Now we will need the second time derivative of the scale factor s:

\begin{align}
\frac{1}{s}\frac{\partial^{2}s}{\partial T_{0}^{2}}=\frac{\partial}{\partial T_{0}}\left(\frac{1}{s}\frac{\partial s}{\partial T_{0}}\right)+\left(\frac{1}{s}\frac{\partial s}{\partial T_{0}}\right)^{2}\,.
\end{align}

We will use the commutator (\ref{eq:commutation rule}) and eqn. (\ref{eq:adot})
to get

\begin{widetext}
\begin{align}
\frac{1}{s}\frac{\partial^{2}s}{\partial T_{0}^{2}} & =\frac{1}{3}\left\langle \frac{\partial_{\mu}T}{\normsq}\,\partial^{\mu}\left(\frac{\theta}{(\normsq)^{1/2}}\right)\right\rangle +\frac{1}{3}\left\langle \frac{\theta^{2}}{\normsq}\right\rangle -\frac{3}{9}\left\langle \frac{\theta}{\norm}\right\rangle ^{2}= \\ \nonumber
& =-\frac{1}{3}\left\langle \frac{\partial_{\mu}T\,\partial^{\mu}\theta}{(\normsq)^{^{3/2}}}\right\rangle +\frac{1}{3}\left\langle \frac{\theta^{2}}{\normsq}\right\rangle -\frac{2}{9}\left\langle \frac{\theta}{(\normsq)^{1/2}}\right\rangle ^{2}-\frac{1}{6}\left\langle \frac{\partial_{\mu}T\partial^{\mu}(\partial_{\nu}T\partial^{\nu}T)}{(\normsq)^{5/2}}\right\rangle \,,
\end{align}
\end{widetext}

rewriting

\begin{align}
\partial^{\mu}T\partial_{\mu}\theta=\nabla_{T}\theta=-(\normsq)^{1/2}n^{\mu}\nabla_{\mu}\theta\,,
\end{align}

we can insert (\ref{eq:Rauchaudhuri}) into the first term

\begin{widetext}
\begin{align}
& \frac{1}{s}\frac{\partial^{2}s}{\partial T_{0}^{2}}=  \frac{1}{3}\left\langle \frac{\theta^{2}}{\normsq}\right\rangle -\frac{2}{9}\left\langle \frac{\theta}{(\normsq)^{1/2}}\right\rangle ^{2}-\frac{1}{6}\left\langle \frac{\partial_{\mu}T\partial^{\mu}(\partial_{\mu}T\partial^{\mu}T)}{(\normsq)^{5/2}}\right\rangle  + \\ \nonumber
& -\frac{1}{3}\left\{ 2\left\langle \frac{\sigma^{2}}{\normsq}\right\rangle +\frac{1}{3}\left\langle \frac{\theta^{2}}{\normsq}\right\rangle -\left\langle \frac{\nabla_{\mu}(n^{\mu}\nabla_{\mu}n_{\nu})}{\normsq}\right\rangle \right\} 
+\frac{8\pi G_{N}}{3}\left\{ \left\langle \frac{\tilde{T}_{\mu\nu}n^{\mu}n^{\nu}+\frac{1}{2}\tilde{T}}{(\normsq)^{1/2}}\right\rangle +\frac{1}{2}\left\langle \frac{\epsilon}{\d_{\m}T\d^{\m}T}+2\epsilon\right\rangle \right\} \,.
\end{align}
\end{widetext}

\vspace{3mm}This is the effective Raychaudhuri equation for the domain.
There is an analogy to the second Friedman equation, since it describes
the acceleration of the domain dependent scale factor s.

\vspace{3mm}

\subsection{Evaluation in the ADM gauge}

In this subsection we will evaluate the general equations in the ADM
gauge, which was introduced above, and compare them to the effective
equations obtained by T. Buchert. As a reminder in the ADM gauge $n_{\mu}=N\,(-1,0,0,0)$
and $n^{\mu}=\frac{1}{N}(1,\ -N^{i})$ where $N$ is the lapse and
$N^{i}$ the shift. Since we have chosen $n_{\mu}$ to be orthogonal to the spatial hypersurfaces
the shift vectors vanish i.e. the metric reads: $ds^{2}=-N^2dt^2+h_{ij}dx^{i}dx^{j}$.
Especially we get for

\begin{align}
\normsq=-\partial_{0}T\, g^{00}\partial_{0}T=\frac{1}{N^{2}}(\partial_{0}T)^{2}\,,
\end{align}

and

\begin{align}
\partial_{0}T_{0}=1\,,
\end{align}

since $T_{0}$ defines the hypersurface of constant physical time.

Given that, the Hamiltonian constraint reduces to

\begin{widetext}
\begin{align}
\left(\frac{1}{s}\frac{\partial s}{\partial T_{0}}\right)^{2} & =-\frac{1}{6}\,\left\langle R_{s}N^{2}\right\rangle -\frac{1}{6}\left\{ \frac{2}{3}\left(\left\langle N^{2}\theta^{2}\right\rangle -\left\langle N\,\theta\right\rangle ^{2}\right)-2\left\langle N^{2}\sigma^{2}\right\rangle \right\} 
+\frac{8\pi G_{N}}{3}\left\langle N^{2}\,\tilde{T}_{\m\n}n^{\m}n^{\n}+\epsilon\right\rangle = \nonumber \\ 
& =-\frac{1}{6}\,\left\langle R_{s}N^{2}\right\rangle -\frac{1}{6}Q_{D}+\frac{8\pi G_{N}}{3}\left\langle N^{2}\,\tilde{T}_{\m\n}n^{\m}n^{\n}+\epsilon\right\rangle  \,.
\end{align}
\end{widetext}

And the Raychaudhuri equation reads as

\begin{widetext}
\begin{align}
-\frac{1}{s}\frac{\partial^{2}s}{\partial T_{0}^{2}} & =-\frac{2}{9}\left(\left\langle N^{2}\theta^{2}\right\rangle -\left\langle N\theta\right\rangle ^{2}\right)+\frac{2}{3}\left\langle N^{2}\sigma^{2}\right\rangle -\frac{1}{3}\left\langle \theta\partial_{0}N\right\rangle -\frac{1}{3}\left\langle N\, h^{ij}\nabla_{i}\nabla_{j}N\right\rangle 
+\frac{8\pi G_{N}}{3}\left\langle N^{2}\left(\tilde{T}_{\mu\nu}n^{\mu}n^{\nu}+\frac{1}{2}\tilde{T}\right)+\epsilon\right\rangle = \nonumber \\ 
& =-\frac{1}{3}Q_{D}-\frac{1}{3}\left\langle \theta\partial_{0}N\right\rangle -\frac{1}{3}\left\langle N\, h^{ij}\nabla_{i}\nabla_{j}N\right\rangle +\frac{8\pi G_{N}}{3}\left\langle \left(\tilde{T}_{\mu\nu}n^{\mu}n^{\nu}+\frac{1}{2}\tilde{T}\right)+\epsilon\right\rangle  \,.
\end{align}
\end{widetext}

Setting $N=1$ we obtain:

The Hamilton constraint density

\begin{widetext}
\begin{align}
\left(\frac{1}{s}\frac{\partial s}{\partial T_{0}}\right)^{2} & =-\frac{1}{6}\,\left\langle R_{s}\right\rangle -\frac{1}{6}\left\{ \frac{2}{3}\left(\left\langle \theta^{2}\right\rangle -\left\langle \theta\right\rangle ^{2}\right)-2\left\langle \sigma^{2}\right\rangle \right\} 
+\frac{8\pi G_{N}}{3}\left\langle \tilde{T}_{\m\n}n^{\m}n^{\n}+\epsilon\right\rangle = \nonumber \\
& =-\frac{1}{6}\,\left\langle R_{s}\right\rangle -\frac{1}{6}Q_{D}+\frac{8\pi G_{N}}{3}\left\langle \tilde{T}_{\m\n}n^{\m}n^{\n}+\epsilon\right\rangle \,.
\end{align}
\end{widetext}

\vspace{3mm}And the Raychaudhuri equation:

\begin{widetext}
\begin{align}
-\frac{1}{s}\frac{\partial^{2}s}{\partial T_{0}^{2}}& =-\frac{2}{9}\left(\left\langle \theta^{2}\right\rangle -\left\langle \theta\right\rangle ^{2}\right)+\frac{2}{3}\left\langle \sigma^{2}\right\rangle 
+\frac{8\pi G_{N}}{3}\left\langle \left(\tilde{T}_{\mu\nu}n^{\mu}n^{\nu}+\frac{1}{2}\tilde{T}\right)+\epsilon\right\rangle  \nonumber \\
& =-\frac{1}{3}Q_{D}+\frac{8\pi G_{N}}{3}\left\langle \left(\tilde{T}_{\mu\nu}n^{\mu}n^{\nu}+\frac{1}{2}\tilde{T}\right)+\epsilon\right\rangle \,.
\end{align}
\end{widetext}

\vspace{3mm}Note: The coordinate dust energy $\epsilon$ has a negative
effect on the acceleration.

The averaging functional with the window function reduces in this
gauge to

\begin{align}
\left\langle S,\Omega\right\rangle =\frac{\int_{\Sigma_{0}}\sqrt{h(x)}S(x)\Theta(r_{0}-B(x))d^{3}x}{\int_{\Sigma_{0}}\sqrt{h(x)}\Theta(r_{0}-B(x))d^{3}x}\,.
\end{align}

\vspace{3mm}

\subsection{The average functional in the ADM gauge}

To perform integration in general we need a push forward from the
manifold into the $\mathcal{\mathbb{R^{\mathrm{n}}}}$, in our case
a map from the dust space into $\mathcal{\mathbb{R^{\mathrm{n}}}}$
is required. Define: $\Psi:=\Phi\circ X\circ Z^{-1}$ s.t. $\Psi:\, S(T)\rightarrow\mathbb{R^{\mathrm{n}}}$
where $\Phi$ is the map $\Phi:\, M\rightarrow\mathbb{R^{\mathrm{n}}}$,
$Z^{-1}$ the map $Z^{-1}:\, S(T)\rightarrow\Sigma$, $X$ the map
$X:\,\Sigma\rightarrow M$ and $S(T)$ the dust-space slice.

\begin{figure}[H]
\begin{centering}
\includegraphics[scale=0.5]{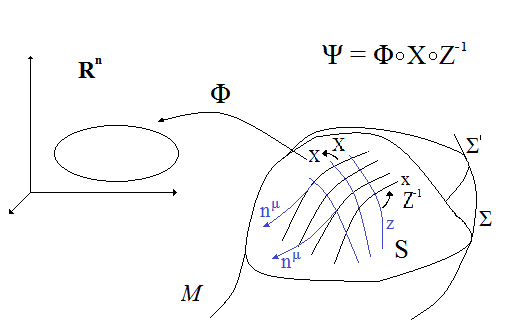}
\par\end{centering}

\caption{The map $\Psi$ used for the push forward into $\mathcal{\mathbb{R^{\mathrm{n}}}}$. This map is a combination of maps from the dust-space slice of constant dust-time to the space of internal variables: $Z^{-1}$, the map from this internal space to the manifold $\mathcal{M}$: $X$ and the map form the manifold to the $\mathcal{\mathbb{R^{\mathrm{n}}}}$: $\Phi$.}
\end{figure}

It was already shown that $Z^{2}$ is gauge invariant so now it is
legitimate to fix a gauge. A legitimate choice would be $Z^{-1} =  \mathbb{I}$
the identity, $X$ the map from the hypersurface coordinates $x^{a}=Z^{i}$
to the manifold coordinates $X^{\m}$ and therefore $S=\Sigma$. In
the ADM gauge the coordinates are $t$ and $x^{a}$ as read of from
the metric $ds^{2}=-dt^{2}+h_{ab}dx^{a}dx^{b}$. This leads to

\begin{align}
Z^{2}(x)=Z^{i}h_{ij}Z^{j}=x^{a}h_{ab}\, x^{b}\,.
\end{align}

Rewriting the average functional one obtains

\begin{align}
\left\langle S,\Omega\right\rangle &  =\frac{\int_{\Sigma_{0}}\sqrt{h(x)}S(x)\Theta(r_{0}-Z^{2}(x))d^{3}x}{\int_{\Sigma_{0}}\sqrt{h(x)}\Theta(r_{0}-Z^{2}(x))d^{3}x} \\ \nonumber
& =\frac{\int_{0}^{r_{0}^{2}}\int_{0}^{2\pi}\int_{0}^{\pi}\, drd\phi d\t \sin(\t)r^{2}\sqrt{h}S(r,\phi,\t)}{\int_{0}^{r_{0}^{2}}\int_{0}^{2\pi}\int_{0}^{\pi}\, drd\phi d\t \sin(\t)r^{2}\sqrt{h}} \,,
\end{align}

which is a volume integral over a sphere of radius $r_{0}^{2}$. 
With this formalism the effect of inhomogeneity on an effective Friedmann model can 
be estimated and an example of application to the ideal fluid cosmology can be found in the Appendix \ref{app:IFC}.

\vspace{3mm}

\subsection{Comparison to Buchert equations}

The difference between our formalism and the averaging applied by
T. Buchert in \cite{Buchert:1999er} is that we did not choose a gauge from the
beginning but deparametrized the manifold with dust. This introduced
a time in the theory. Furthermore we found a way to perform gauge
invariant averages over finite volumes. Hence the averaged quantities
$\left\{ s\,\dot{s}\,\ddot{s}\,\left\langle \epsilon\right\rangle \right\} $
became physical observables and we obtained equations governing their
dynamics. The equations we get after fixing the ADM gauge are slightly
different to the ones in \cite{Buchert:1999er}. We recover Buchert's
equations in our general formalism if we set the rest energy momentum
$\tilde{T}_{\m\n}$ to zero. Therefore one could think that in Buchert's
formalism just the geometry is averaged. The problem with this point
of view is that even an infinitesimal energy density $\epsilon$ leads
to a geometry arbitrary far away from the vacuum geometry, since there
is no smooth limit $\epsilon\rightarrow0$ .

Hence one has to see that Buchert's equations describe correctly the evolution of a cosmology
filled with one type of pressure-free fluid in the rest frame of this
fluid. We see that in this special case the procedures of gauge fixing
and averaging commute. Our formalism allows to generalize the averaging
to an almost arbitrary energy momentum tensor. The only restriction
is that the energy density must not vanish anywhere. Especially it
is not possible to apply this formalism to space-time devoid of all matter.

\vspace{3mm}

\section{Summary}

A manifestly gauge invariant averaging formalism for finite volumes
was presented in this work. The  question of gauge
dependence of the backreaction was resolved. The model used just required
the assumption of a non-vanishing energy density. In this case a dust
energy momentum tensor can be separated from the total energy density.
The coordinate fields of the dust were used as a reference frame to
perform averaging. 

The derived equations for the average of an arbitrary energy momentum
tensor were evaluated in the ADM gauge, which corresponds to choosing
the rest frame of the dust. Furthermore, as an example the equation
is evaluated for an energy momentum tensor of an ideal fluid. The
resulting equation was also studied in the ADM gauge. An interesting
phenomenon was observed, as discussed in the ideal fluid cosmology in the Appendix \ref{app:IFC}.
It appears that ``tilt'' effects (i.e.
non-co-linearity of the dust flow and the fluid's flow) contribute
negatively to the acceleration of the domain dependent scale factor. 

The derived equations in the ADM gauge were compared to the Buchert
equations. We are able to interpret the Buchert equations in the
gauge independent framework. It turned out that they describe the
averaged behaviour of a pressure-free fluid, using this fluid itself
as a reference and choosing it's rest frame. It turns out that in
this special case gauge fixing and averaging commute. 

By deriving gauge independent equations for the cosmological backreaction,
we proved that it is in principle an observable. With this formalism at hand it is possible
to study cosmological data and can trust the derived equations to
correctly describe the observations, once we identify a system of freely falling observers.

One could go one step further and argue that given the necessity to explain the abundance of a matter species, which mainly -or entirely- interacts 
gravitationally, we can view our system with the coordinate dust as a realistic cosmological model. In which 
case the dust is the Dark Matter which behaves, according to observations as a pressure-less fluid.

An interesting and may be fundamental observation is, that it is not possible to define 
averaged observables in a space-time devoid of matter.

\section*{Acknowledgments}

I would like to thank K. Giesel, S. Meyer  and M. Bartelmann for very helpful discussions.  
I acknowledge support from the IMPRS for Precision Tests of Fundamental Symmetries.
Most warmly I would like to thank S. Hofmann who has suggested to study this topic and 
has provided vital guidance during the completion of this work. 
\appendix
\section{\label{app:IFC}Example of an Ideal fluid cosmology}

A cosmology equipped with an energy momentum tensor of an ideal fluid
will be studied. This is a reasonable choice for the energy momentum,
since it is the most general tensor fulfilling the symmetries of an
isotropic universe and since we know from the CMB that the universe
has been extremely isotropic this model is a reasonable choice. This
procedure can be easily generalized to an arbitrary number of ideal
non interacting fluids.

\vspace{3mm}

\subsubsection*{Normal projection of the equations of motion}

The Einstein tensor multiplied with the time flow vectors (which is
equivalent to the Hamiltonian constraint density) gives:

\begin{align}
G^{\mu\nu}n_{\mu}n_{\nu}=\frac{1}{2}R_{s}+\frac{1}{3}\theta^2-\sigma^2 \,,
\end{align}

where the expansion rate and the shear are defined w.r.t $n_{\mu}$,
the $R_{s}$ is the spatial curvature associated with the projector
on the hypersurface orthogonal to $n_{\mu}$. On the other hand one
has from the matter contribution

\begin{align}
T^{\mu\nu}n_{\mu}n_{\nu} =(T_{\textrm{fluid}}^{\mu\nu}+T_{\textrm{dust}}^{\mu\nu})n_{\mu}n_{\nu}\,,
\end{align}

with

\begin{align}
T_{\textrm{fluid}}^{\mu\nu}n_{\mu}n_{\nu}& =(\rho+p)(u^{\mu}n_{\mu})^2-p \nonumber \\ 
& =\rho+(\rho+p)\, \sinh^2(\alpha_{T})
\end{align}

and

\begin{align}
T_{\textrm{dust}}^{\mu\nu}n_{\mu}n_{\nu} & =\epsilon(v_{\mu}n^{\mu})^2=\epsilon(-\partial_{\mu}Tn^{\mu})^2 \nonumber \\
& =\epsilon(-\partial_{\mu}T\partial^{\mu}T) \,,
\end{align}

where the tilt angle $\alpha_{T}$ defined via $\sinh^2(\alpha_{T}):=(u^{\mu}n_{\mu})^2-1$.
Furthermore as discussed above $v^{\mu}=\norm\: n^{\mu}$ since we
said that the velocity field $W^{k}$ vanishes. Now the Hamiltonian
constraint can be averaged. For normalization divide it first by $-3\,\partial_{\mu}T\partial^{\mu}T$
and apply the averaging functional introduced above

\begin{widetext}
\begin{align}
& \left\langle \frac{1}{6}\frac{R_{s}}{\normsq}+\frac{1}{9}\frac{\theta^2}{\normsq}-\frac{1}{3}\frac{\sigma^2}{\normsq}\right\rangle +\frac{1}{9}\left\langle \frac{\theta}{\norm}\right\rangle ^{2}-\frac{1}{9}\left\langle \frac{\theta}{\norm}\right\rangle ^{2}=\label{eq:hamilton constraint-1} \nonumber \\ 
& =\frac{8\pi G_{N}}{3}\left\langle \frac{\rho+(\rho+p)\sinh^2(\alpha_{T})}{\normsq}+\epsilon\right\rangle \,.
\end{align}
\end{widetext}

As above we rewrite (\ref{eq:hamilton constraint-1}) as

\begin{widetext}
\begin{align}
\left(\frac{1}{s}\frac{\partial s}{\partial T_{0}}\right)^{2}=-\frac{1}{6}\,\left\langle \frac{R_{s}}{\normsq}\right\rangle -\frac{1}{6}Q_{D}+\frac{8\pi G_{N}}{3}\left\langle \frac{\rho+(\rho+p) \sinh^2(\alpha_{T})}{\normsq}+\epsilon\right\rangle 
\end{align}
\end{widetext}

\vspace{3mm}This is the effective equation describing the velocity
of the expansion of our domain-dependent scale factor. With the definitions
as above we can write the cosmological balance equation as:

\begin{align}
H_{D}:=\frac{1}{s}\frac{\partial s}{\partial T_{0}}\,,
\end{align}

\begin{align}
\Omega_{m}:=\frac{8\pi G_{N}}{3\, H_{D}^{2}}\left\langle \frac{\rho+(\rho+p)\sinh^2(\alpha_{T})}{\normsq}\right\rangle \,,
\end{align}

\begin{align}
\Omega_{\epsilon}:=\frac{8\pi G_{N}}{3\, H_{D}^{2}}\,\left\langle \epsilon\right\rangle \,,
\end{align}

\begin{align}
\Omega_{k}:=-\frac{\left\langle R\right\rangle }{6H_{D}^{2}}\,,
\end{align}

\begin{align}
\Omega_{Q}:=-\frac{Q_{D}}{6H_{D}^{2}}\,,
\end{align}

\begin{align}
\Omega_{k}+\Omega_{Q}+\Omega_{\epsilon}+\Omega_{m}=1\,.
\end{align}

From the energy balance equation we see that the backreaction term
$\Omega_{Q}$ opens the possibility of contributing to an effective
cosmological constant term in this model, which is in principle close
to $\Lambda$-CDM.

\vspace{3mm}

\subsubsection*{Raychaudhuri equation:}

Combining the Hamiltonian constraint with the trace of the Einstein
equations projected orthogonal to the time flow, one gets:

\begin{align}
R_{\mu\nu}n^{\mu}n^{\nu}=T_{\mu\nu}h^{\mu\nu}-\frac{1}{2}T
\end{align}

Which is equivalent to

\begin{align}
& -n^{\mu}\nabla_{\mu}\theta=2\sigma^2+\frac{1}{3}\theta^2-\nabla^{\nu}(n^{\mu}\nabla_{\mu}n_{\nu}) \nonumber \\& +(T_{\mu\nu}-\frac{1}{2}g_{\mu\nu}T)n^{\mu}n^{\nu} \,.
\label{eq:Rauchaudhuri-1}
\end{align}

In our model the last term gives

\begin{align}
& (T_{\mu\nu}-\frac{1}{2}g_{\mu\nu}T)n^{\mu}n^{\nu}  =T_{\mu\nu}^{\textrm{fluid}}n^{\mu}n^{\nu}+T_{\mu\nu}^{\textrm{dust}}n^{\mu}n^{\nu}+\frac{1}{2}T  \\  \nonumber
& =\rho+\sinh^{2}(\alpha_{T})(\rho+p)+\epsilon(-\partial_{\mu}T\partial^{\mu}T)+\frac{1}{2}(3p-\rho-\epsilon)\,.
\end{align}

Again we will use the commutator (\ref{eq:commutation rule}) and
Eqn. (\ref{eq:adot}) to get

\begin{widetext}
\begin{align}
& \frac{1}{s}\frac{\partial^{2}s}{\partial T_{0}^{2}}=\frac{1}{3}\left\langle \frac{\partial_{\mu}T}{\normsq}\,\partial^{\mu}\left(\frac{\theta}{(\normsq)^{1/2}}\right)\right\rangle +\frac{1}{3}\left\langle \frac{\theta^{2}}{\normsq}\right\rangle -\frac{3}{9}\left\langle \frac{\theta}{\norm}\right\rangle ^{2}= \nonumber \\
& =-\frac{1}{3}\left\langle \frac{\partial_{\mu}T\,\partial^{\mu}\theta}{(\normsq)^{^{3/2}}}\right\rangle +\frac{1}{3}\left\langle \frac{\theta^{2}}{\normsq}\right\rangle -\frac{2}{9}\left\langle \frac{\theta}{(\normsq)^{1/2}}\right\rangle ^{2}-\frac{1}{6}\left\langle \frac{\partial_{\mu}T\partial^{\mu}(\partial_{\nu}T\partial^{\nu}T)}{(\normsq)^{5/2}}\right\rangle \,.
\end{align}
\end{widetext}

Rewriting

\begin{align}
\partial^{\mu}T\partial_{\mu}\theta=\nabla_{T}\theta=-(\normsq)^{1/2}n^{\mu}\nabla_{\mu}\theta\,,
\end{align}

we can insert (\ref{eq:Rauchaudhuri-1}) into the first term

\begin{widetext}
\begin{align}
& \frac{1}{s}\frac{\partial^{2}s}{\partial T_{0}^{2}}=\frac{1}{3}\left\langle \frac{\theta^{2}}{\normsq}\right\rangle -\frac{2}{9}\left\langle \frac{\theta}{(\normsq)^{1/2}}\right\rangle ^{2}-\frac{1}{6}\left\langle \frac{\partial_{\mu}T\partial^{\mu}(\partial_{\mu}T\partial^{\mu}T)}{(\normsq)^{5/2}}\right\rangle +-\frac{1}{3}\left\{ 2\left\langle \frac{\sigma^{2}}{\normsq}\right\rangle \right. \\ \nonumber
& \left. +\frac{1}{3}\left\langle \frac{\theta^{2}}{\normsq}\right\rangle -\left\langle \frac{\nabla_{\mu}(n^{\mu}\nabla_{\mu}n_{\nu})}{\normsq}\right\rangle \right\}  
+\frac{8\pi G_{N}}{3}\left\{ \left\langle \frac{(\rho+p) \sinh^{2}(\alpha_{T})+\rho}{(\normsq)^{1/2}}\right\rangle +\frac{1}{2}\left\langle \frac{3p-\rho-\epsilon}{\normsq}+2\epsilon\right\rangle \right\} \,.
\end{align}
\end{widetext}

\vspace{3mm}This is the effective Raychaudhuri equation for the domain
in the case of the energy momentum tensor being that of an ideal fluid. 

\vspace{3mm}

\subsection{Evaluation in the ADM gauge}

In this subsection we will evaluate the equations of 4.2.5 in the
ADM gauge, which was introduced above. 

The Hamiltonian constraint density reduces to

\begin{widetext}
\begin{align}
\left(\frac{1}{s}\frac{\partial s}{\partial T_{0}}\right)^{2} & =-\frac{1}{6}\,\left\langle R_{s}N^{2}\right\rangle -\frac{1}{6}\left\{ \frac{2}{3}\left(\left\langle N^{2}\theta^{2}\right\rangle -\left\langle N\,\theta\right\rangle ^{2}\right)-2\left\langle N^{2}\sigma^{2}\right\rangle \right\} 
+\frac{8\pi G_{N}}{3}\left\langle N^{2}\,(\rho+(\rho+p)\sinh^2(\alpha_{T}))+\epsilon\right\rangle =
\nonumber \\
& =-\frac{1}{6}\,\left\langle R_{s}N^{2}\right\rangle -\frac{1}{6}Q_{D}+\frac{8\pi G_{N}}{3}\left\langle N^{2}\,(\rho+(\rho+p) \sinh^2(\alpha_{T}))+\epsilon\right\rangle \,.
\end{align}
\end{widetext}

And the Raychaudhuri equation to

\begin{widetext}
\begin{align}
 -\frac{1}{s}\frac{\partial^{2}s}{\partial T_{0}^{2}} & =-\frac{2}{9}\left(\left\langle N^{2}\theta^{2}\right\rangle -\left\langle N\theta\right\rangle ^{2}\right)+\frac{2}{3}\left\langle N^{2}\sigma^{2}\right\rangle -\frac{1}{3}\left\langle \theta\partial_{0}N\right\rangle -\frac{1}{3}\left\langle N\, h^{ij}\nabla_{i}\nabla_{j}N\right\rangle + \nonumber \\ 
& +\frac{4\pi G_{N}}{3}\left\langle 2\, N^{2}((\rho+p) \sinh^{2}(\alpha_{T}))+N^{2}(\rho+3p)+2\epsilon\right\rangle  \\ \nonumber
& =-\frac{1}{3}Q_{D}-\frac{1}{3}\left\langle \theta\partial_{0}N\right\rangle -\frac{1}{3}\left\langle N\, h^{ij}\nabla_{i}\nabla_{j}N\right\rangle +\frac{4\pi G_{N}}{3}\left\langle 2\, N^{2}((\rho+p) \sinh^{2}(\alpha_{T}))+N^{2}(\rho+3p)+2\epsilon\right\rangle \,.
\end{align}
\end{widetext}

Setting $N=1$ we obtain:

The Hamilton constraint density

\begin{widetext}
\begin{align}
 \left(\frac{1}{s}\frac{\partial s}{\partial T_{0}}\right)^{2} & =-\frac{1}{6}\,\left\langle R_{s}\right\rangle -\frac{1}{6}\left\{ \frac{2}{3}\left(\left\langle \theta^{2}\right\rangle -\left\langle \theta\right\rangle ^{2}\right)-2\left\langle \sigma^{2}\right\rangle \right\} 
+\frac{8\pi G_{N}}{3}\left\langle (\rho+(\rho+p) \sinh^2(\alpha_{T}))+\epsilon\right\rangle  \nonumber \\
& =-\frac{1}{6}\,\left\langle R_{s}\right\rangle -\frac{1}{6}Q_{D}+\frac{8\pi G_{N}}{3}\left\langle \rho+(\rho+p) \sinh^2(\alpha_{T})+\epsilon\right\rangle \,.
\end{align}
\end{widetext}

And the Raychaudhuri equation

\begin{widetext}
\begin{align}
 -\frac{1}{s}\frac{\partial^{2}s}{\partial T_{0}^{2}} & =-\frac{2}{9}\left(\left\langle \theta^{2}\right\rangle -\left\langle \theta\right\rangle ^{2}\right)+\frac{2}{3}\left\langle \sigma^{2}\right\rangle +\frac{8\pi G_{N}}{3}\left\langle 2\,((\rho+p) \sinh^{2}(\alpha_{T}))+(\rho+3p)+2\epsilon\right\rangle =
\\ \nonumber
& =\frac{8\pi G_{N}}{3}\left\langle 2\,((\rho+p) \sinh^{2}(\alpha_{T}))+(\rho+3p)+2\epsilon\right\rangle -\frac{1}{3}Q_{D}\,.
\end{align}
\end{widetext}

Observations:

\begin{itemize}
\item Both the Hamiltonian constraint and the Raychaudhuri equation contain
the energy density of the coordinate dust. In the second equation
it contributes negatively to the acceleration of the scale factor.
\item The fact that generically the flow of the coordinate dust is non parallel
to the flow of the fluid which is averaged, is manifested in the $\sinh^{2}(\alpha_{T})$
term. This as well contributes negatively to the acceleration of the
scale factor $s$.
\item After the gauge choice the spatial integral is an ordinary 3-dimensional
integral.

\item The backreaction $Q_{D}$ has the potential to contribute to the acceleration.
This has been addressed in the literature \cite{Rasanen:2011ki,Li:2008yj,Buchert:2011yu} and certainly deserves attention.
\end{itemize}


\end{document}